\documentclass[12pt,preprint]{aastex}

\def\ltsima{$\; \buildrel < \over \sim \;$}
\def\gtsima{$\; \buildrel > \over \sim \;$}
\def\lsim{\lower.5ex\hbox{\ltsima}}
\def\gsim{\lower.5ex\hbox{\gtsima}}
\def\lapp{\ifmmode\stackrel{<}{_{\sim}}\else$\stackrel{<}{_{\sim}}$\fi}
\def\gapp{\ifmmode\stackrel{>}{_{\sim}}\else$\stackrel{<}{_{\sim}}$\fi}

\newdimen\minuswidth    
\setbox0=\hbox{$-$}
\minuswidth=\wd0
\catcode`@=\active
\def@{\kern\minuswidth}
\setbox0=\hbox{\rm0}
 
\shorttitle{The binary content of M10} 
\shortauthors{Dalessandro et al.}
 
\begin{document} 
\title{The binary fraction in the globular cluster M10 (NGC 6254):
  comparing core and outer regions \footnote{Based on observations
    collected with the NASA/ESA {\it HST}, obtained at the Space
    Telescope Science Institute, which is operated by AURA, Inc.,
    under NASA contract NAS5-26555.}  }

\author{
E. Dalessandro\altaffilmark{2},
B. Lanzoni\altaffilmark{2}, 
G. Beccari\altaffilmark{3}
A. Sollima\altaffilmark{4},
F.R. Ferraro\altaffilmark{2},
and M. Pasquato\altaffilmark{2}
}
\affil{\altaffilmark{2} Dipartimento di Astronomia, Universit\`a degli Studi
di Bologna, via Ranzani 1, I--40127 Bologna, Italy}
\affil{\altaffilmark{3} ESO - European Southern Observatory,
  Karl-Swarzschild Str. 2, D-85748 Garching bei M\"unchen,  Germany} 
\affil{\altaffilmark{4} INAF Osservatorio Astronomico di Padova,
  Vicolo dellOsservatorio 5, I--35122 Padova, Italy}
\date{24 August, 2011}

\begin{abstract}
We study the binary fraction of the globular cluster M10 (NGC~6254)
as a function of radius from the cluster core to the outskirts, by means of a quantitative
analysis of the color distribution of stars relative to the fiducial main sequence.
By taking advantage of two data-sets, acquired with
the Advanced Camera for Survey and the Wide Field Planetary Camera 2
on board the Hubble Space Telescope, we have studied both the core and
the external regions of the cluster. The binary fraction is
found to decrease from $\sim 14\%$ within the core, to $\sim 1.5\%$ in
a region between 1 and 2 half-mass radii from the cluster centre. Such
a trend and the derived values are in agreement with previous results
obtained in clusters of comparable total magnitude. The estimated
binary fraction is sufficient to account for the suppression of mass
segregation observed in M10, without any need to invoke the presence
of an intermediate-mass black hole in its centre.
\end{abstract}
 

\section{INTRODUCTION}
The binary fraction is an essential component in the formation and evolution 
of dynamically active systems, like
globular clusters (GCs). In such dense environments, where stellar
gravitational interactions are very frequent, binaries can exert a
significant influence on both the dynamical evolution of the system
and the properties of its stellar populations.  

Being, on average, more massive than the other stars, binaries tend to
sink into the highly crowded cluster centers, because of equipartition.
 The characteristic timescale of this process (the
relaxation time) depends on the cluster structure and can be even
longer than a Hubble time in the outskirts. Hence, in the outer
regions of GCs we essentially expect to observe primordial binary
systems, i.e., binaries created as part of the star formation process
and evolving undisturbed. In the cluster core, on the other hand, a variety of
dynamical processes (exchange interactions, three-body encounters,
tidal captures, etc.)  can take place, with competing effects on the
binary population: binaries can be destroyed, created or just modified
\citep[e.g.][]{hut92}, with relative efficiencies that still are a
matter of debate in the literature \citep[e.g.][]{ivanova05, hurl07,
  sol08, freg09}.  In general, however, the gravitational encounters
occurring in the core tend to make the binaries harder (more tightly
bound), thus providing a central energy source able to slow down the
cluster core collapse \citep{goodhut89}. Following the N-body
simulations of \citet{gill08}, this energy source could also suppress
the mass segregation process, with a detectable effect on the radial
behavior of the mass function of main-sequence (MS) stars. Since the
same effect could alternatively be due to a central intermediate-mass
black hole \citep[IMBH; see also][herefater B10]{pasqua09, bec10},
if we can measure the fraction of  binaries, then we can say
whether or not we need an IMBH to explain the low level of mass 
segregation that has been observed. Hence, the empirical estimate of
the binary fraction in a sample of GCs representative of different
environments is a prime ingredient for dynamical models, which help us
understand the internal cluster dynamics.

The knowledge of the binary fraction is also crucial for understanding
the properties of puzzling objects like blue stragglers, millisecond
pulsars and cataclysmic variables, which are all thought to be the
by-products of binary evolution \citep[e.g.,][and references
  therein]{mcrea64,romani87,fe01,leigh11}. In particular, the analysis
of the bimodal radial distribution of blue stragglers observed in a
number of GCs \citep[e.g., Ferraro et al. 1997, 2004;][]{ema_6388}
suggests that a non-negligible fraction of these stars is generated by
primordial binaries, which still orbit in isolation in the cluster
outskirts and produce the observed rising branch of the distribution
(Mapelli et al. 2004, 2006; Lanzoni et al. 2007a,b).  The
interpretation of the double blue straggler sequence recently
discovered in the core of M30 also requires a significant fraction of
primordial binaries \citep{fe09_m30}.

Despite its implications, however, the binary fraction in GCs 
still remains badly constrained, because of the challenging observational
requirements.  The main techniques commonly used for its estimate are:
radial velocity variability surveys \citep[e.g.][]{pry89,lath96,
  albr01}, searches for eclipsing binaries \citep[e.g.][]{mateo96,
  cote96}, and the study of the distribution of stars along the
cluster MS in color-magnitude diagrams \citep[CMDs;
  e.g.,][]{romani91,bolte92,rubai97,bell02,clark04,zhao05,sol07,milone08}.
The first two methods rely on the detection of individual binary
systems in a given range of periods and mass ratios.  Hence, the
nature of these methods leads to intrinsic observational biases and a
low detection efficiency.  The latter approach relies on the simple fact
that, since the flux of unresolved binaries is equal to the sum
of the fluxes of the two components, the binaries composed by MS
companions are shifted towards brighter magnitudes with respect to the
single-star MS.  This technique has the advantage of being more
efficient and detecting binary systems regardless of their orbital
periods and inclinations.

For the present paper we used this latter technique to estimate
the binary fraction in the core and the outskirts of M10 (NGC 6254).
This is an ``ordinary'', dynamically relaxed GC, with absolute visual
magnitude $M_V=-7.48$ \citep[][2010 edition]{harris96}, central mass
density $\log\rho_0=3.8$ \citep[$\rho_0$ being in units of
  $M_\odot/$pc$^3$;][]{prymey93}, and half-mass relaxation time
$t_h\sim 0.8$ Gyr \citep[Harris 1996; see also][]{mcLvdM05}.  
The deep and high-quality photometry that B10 obtained for
both the centre and beyond the half-mass radius, allowed them
to study the cluster mass function
at different radial distances.  The resulting mass-segregation profile
is moderately flattened and can be explained by the presence
of either an IMBH of $\sim 10^3 M_\odot$, or a population of binaries
with an initial fraction of 3-5\% (B10). Hence, within the framework
proposed by \citet{gill08}, any empirical constraint on the binary
content in this system would allow us to assess the possible presence of
a central IMBH. In addition, it will provide precious clues 
and constraints that would be useful for a robust interpretation of 
the properties the blue-straggler-population in this cluster
(Emanuele Dalessandro et al. 2011 in preparation).

The paper is organized as follows.  The used data-sets are presented
in Section 2. The method adopted to estimate the binary fraction is
outlined in Section 3.  The results and the discussion are presented
in Section 4 and 5.

\section{THE DATA} \label{data} 
\begin{figure}
\begin{center}
\includegraphics[scale=0.9]{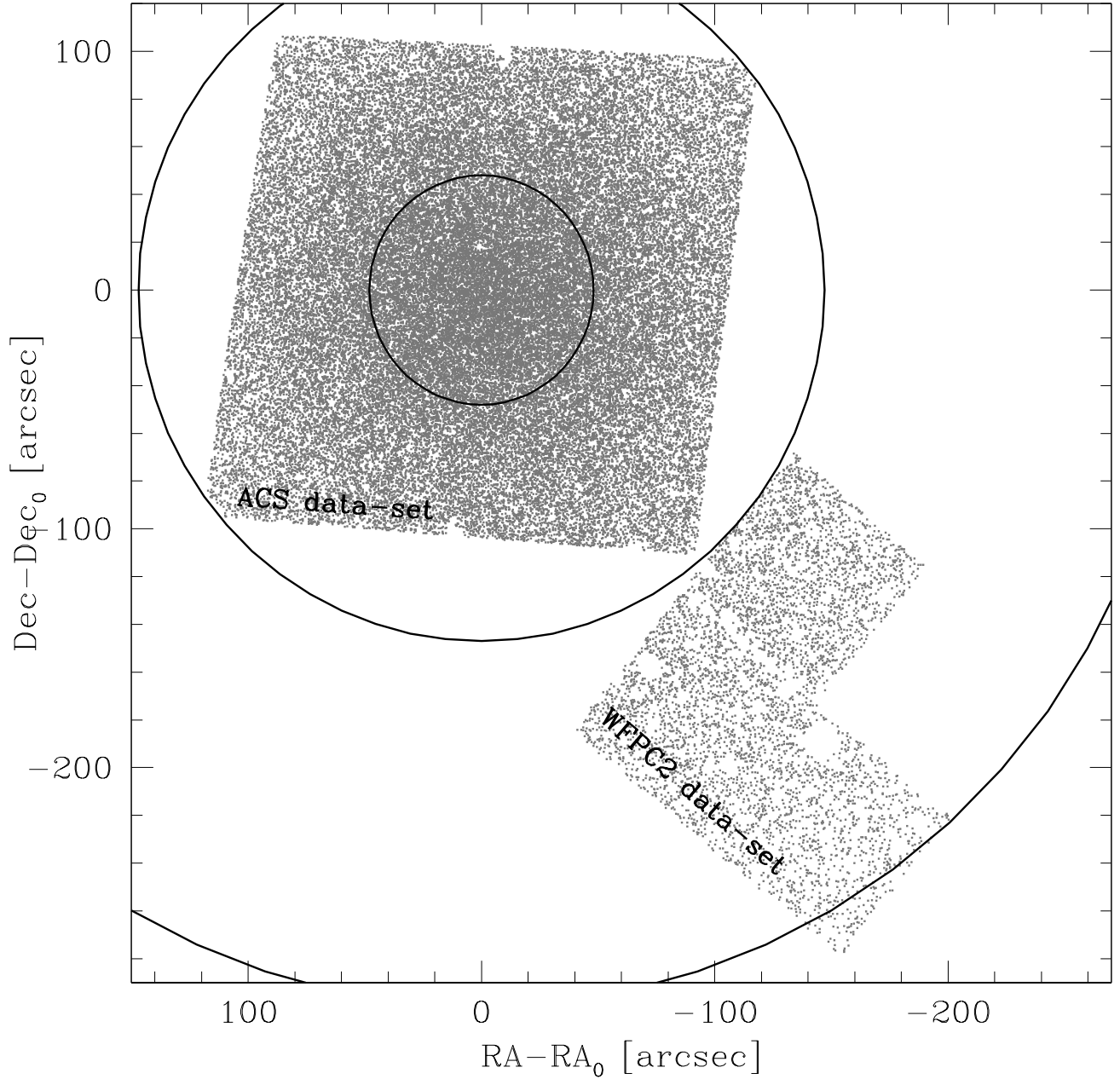}
\caption{Map of the ACS and WFPC2 data-sets. The two circles mark the
  position of the core and half-mass radii ($r_c=48\arcsec$,
  $r_h=147\arcsec$).  The ACS data-set samples the inner portion of
  M10 out to $\sim r_h$, while the WFPC2 data-set (consisting of the
  data acquired with the three wide field cameras) covers a region
  between one and two half-mass radii.}
\label{map}
\end{center}
\end{figure}

The data-set used in the present work (the same as in B10) consists of
a sample of $4\times90sec$ images acquired in the F606W ($V$) and 
$4\times90sec$ images in F814W
filters obtained with the Advanced Camera for Surveys (ACS; GO-10775, PI: Sarajedini),  
complemented 
with $2\times1100sec$ and $2\times1200sec$ images in F606W and
 $2\times1100sec$ and $2\times1200sec$ images obtained with the
 Field Planetary Camera 2 (WFPC2; GO-6113, PI: Paresce) on board the Hubble Space Telescope.
The ACS data-set samples the
cluster central regions, while the WFPC2 one covers an area located
between one and two half-mass radii (see Figure \ref{map}).  The
detailed description of the data reduction, photometric
calibration\footnote{The instrumental $V$ and $I$ magnitudes have been
 calibrated to the VEGAMAG system by following the
  prescription of \citet{sirianni05} and \citet{holtz95} for the ACS
  and the WFPC2 samples, respectively.}  and astrometric solution
procedures is given in B10 and the sample used in the present analysis
contains the same ``bona-fide'' stars selected on the basis of the
quality of the point-spread-function fitting \citep[as measured by the
  DAOPHOTII sharpness parameter;][]{stet87}.  The CMDs of the two
data-sets are shown in Figure \ref{cmd}.  Stars brighter than $I=16$
and $I=19.5$ are saturated in the ACS and WFPC2 samples, respectively.
B10 also performed a detailed photometric completeness study, based on
artificial star experiments, where artificial stars were added to the
original FLT frames and the whole data reduction procedure was repeated.
From the resulting catalogue, listing the input and output positions
and magnitudes for more than 500,000 artificial stars, B10 estimated
that the photometric completeness drops below 50\% at $I\sim 22.5$ in
the innermost region of the cluster, and at $I\sim 25$ for the WFPC2
data-set.

\section{THE ANALYSIS}\label{analysis}
In order to estimate the binary fraction of M10, we followed the
method extensively described in \citet{bell02} and \citet[][hereafter
  S07]{sol07}.  The basic idea is that the magnitude of a binary
system corresponds to the luminosity of the primary (more massive)
star, increased by the contribution of the companion of an amount that
depends on the mass ratio of the two components ($q=M_2/M_1$).  In
fact, since the stars along the MS follow a mass-luminosity relation,
the luminosity of a binary can be written in terms of the mass ratio
of the two components. By definition, $0<q\le 1$ and for $q=1$
(equal-mass binary) the system appears $\sim 0.75$ magnitudes brighter
than the single component, while the luminosity enhancement decreases
for decreasing $q$.  The spanning of all the possible values of $q$,
at different magnitudes of the primary component, produces a
broadening of the single-star MS, to its bright- and red-hand side.
In principle, the ratio between the number of stars lying on the red side 
of the single-star MS and the total number of stars observed along the
``broadened MS'' provides the cluster binary fraction. In practice,
depending on the photometric error of the data, a minimum value of the
mass ratio ($q_{min}$) exists below which it is impossible to
observationally distinguish a binary system from a single MS star.
Moreover, it is necessary to take into account a number of effects,
like stellar blends and the contamination by foreground/background
field stars, which can add spurious sources in the CMD.

Indeed, chance superpositions of two stars (blends) can produce a
luminosity enhancement that mimics the magnitude shift characteristic
of a genuine binary system.  In order to correct for this effect we
analyzed the distribution of the residuals between the input and the
output magnitudes of the artificial star catalogue built by B10 for
the completness study (see previous section). From the asymmetry of
the distribution (which is skewed toward brighter output magnitudes
because of the blending between artificial and real stars) we
estimated that the percentage of blended sources that would mimick 
binary systems with $q>q_{min}$, varies from $\sim
6\%$ in the core, to less than $0.2\%$ in the external regions.

B10 also estimated the Galactic field contamination in the direction
of M10, finding that it is very low: even in the worst case (the WFPC2
data-set), where the number of cluster sources is small, the field stars are
just $\sim 3\%$ of the total sample.  Despite such a low value, for a
proper measurement of the binary fraction we performed a detailed
study of the field contamination as a function of the magnitude.  From
the Galaxy model of \citet{robin03}\footnote{publicly available at
  http://model.obs-besancon.fr/} we retrieved a catalogue covering an
area of $0.5\deg^2$ in the direction of M10, and we randomly extracted
two sub-samples of synthetic stars, scaled to the fields of view of
the ACS and WFPC2 data-sets. Their magnitudes were converted from the
Johnson to the VEGAMAG photometric system adopting the prescriptions
of \citet{sirianni05}. Finally, by exploiting the artificial-star
catalogue used for the completeness study\footnote{For each considered
  synthetic field star, we randomly extracted an artificial object
  with similar magnitude ($\Delta I<0.1$) and we assigned the shifts
  between its input and output magnitudes to the field star, in a effort to
  mimick and take into account the effects of completeness and
  blending.}  (Sect. \ref{data}), we obtained a catalogue of synthetic
field stars that includes the observational biases (incompleteness and
blending), for both the ACS and WFPC2 data-sets.

Once all the contaminant effects are taken into account, the binary
fraction was estimated as the number of stars in the
\emph{``binary population''} divided by the total number of stars, i.e.,
binaries plus genuine, single, MS stars (hereafter the \emph{``MS
  population''}).  The ``MS population'' is defined as the set of stars
having a color difference from the MS ridge line (MSRL) smaller than
three times the typical photometric error at that magnitude level (see
Figure \ref{regions}). The operational definition of the ``binary
population'' is given in Sects. \ref{fmin} and \ref{fglob}.

\section{RESULTS}
\begin{figure}
\begin{center}
\includegraphics[scale=0.9]{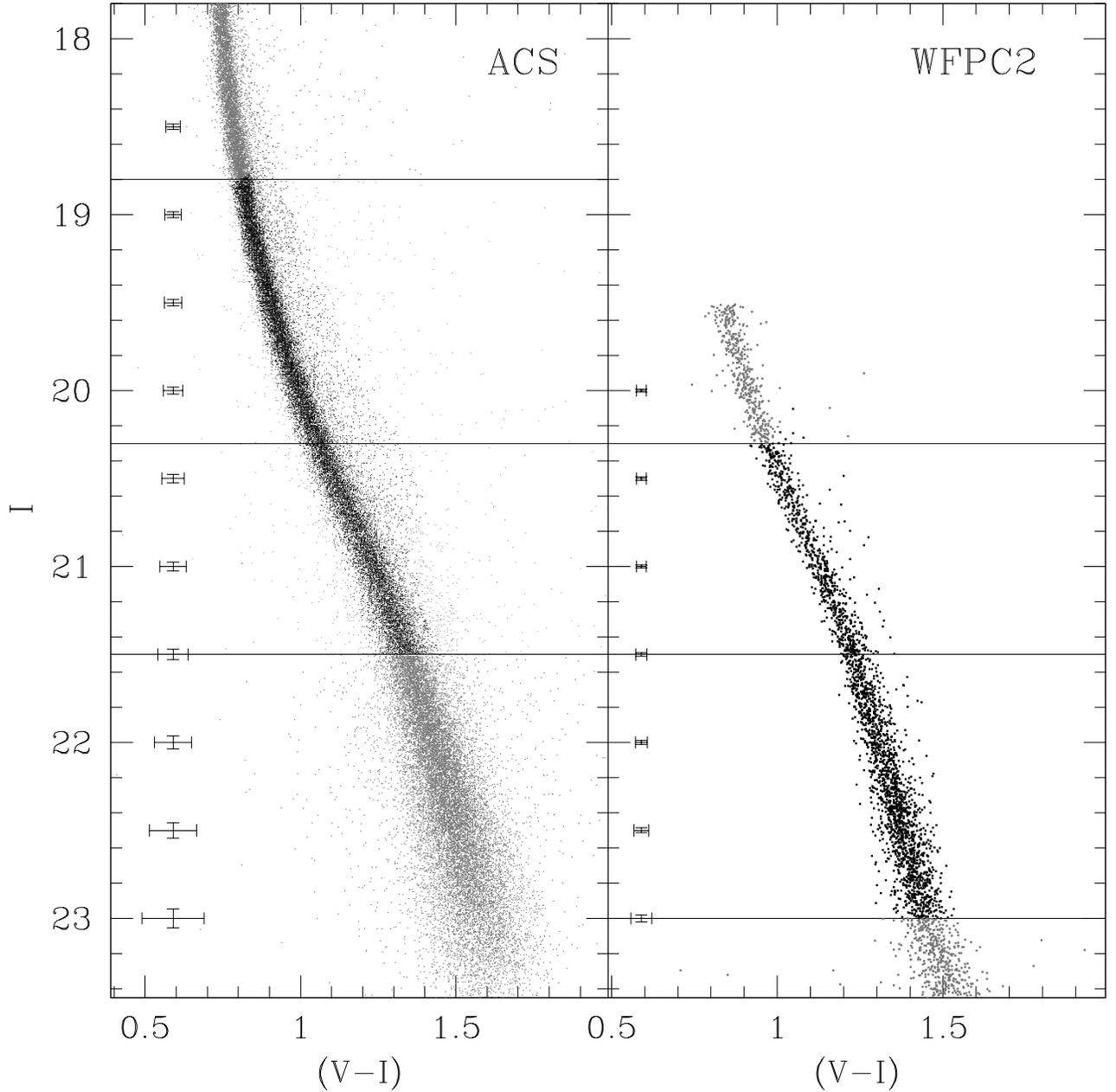}
\caption{CMDs of the ACS and WFPC2 data-sets. The stars used to
  estimate the binary fraction are plotted in black. The three
  horizontal lines mark the values along the MSRL that separate the
  magnitude ranges considered in the work (see
  Sect. \ref{analysis}). The photometric errors at different magnitude
  levels are shown. }
\label{cmd}
\end{center}
\end{figure}

\begin{figure}
\begin{center}
\includegraphics[scale=0.9]{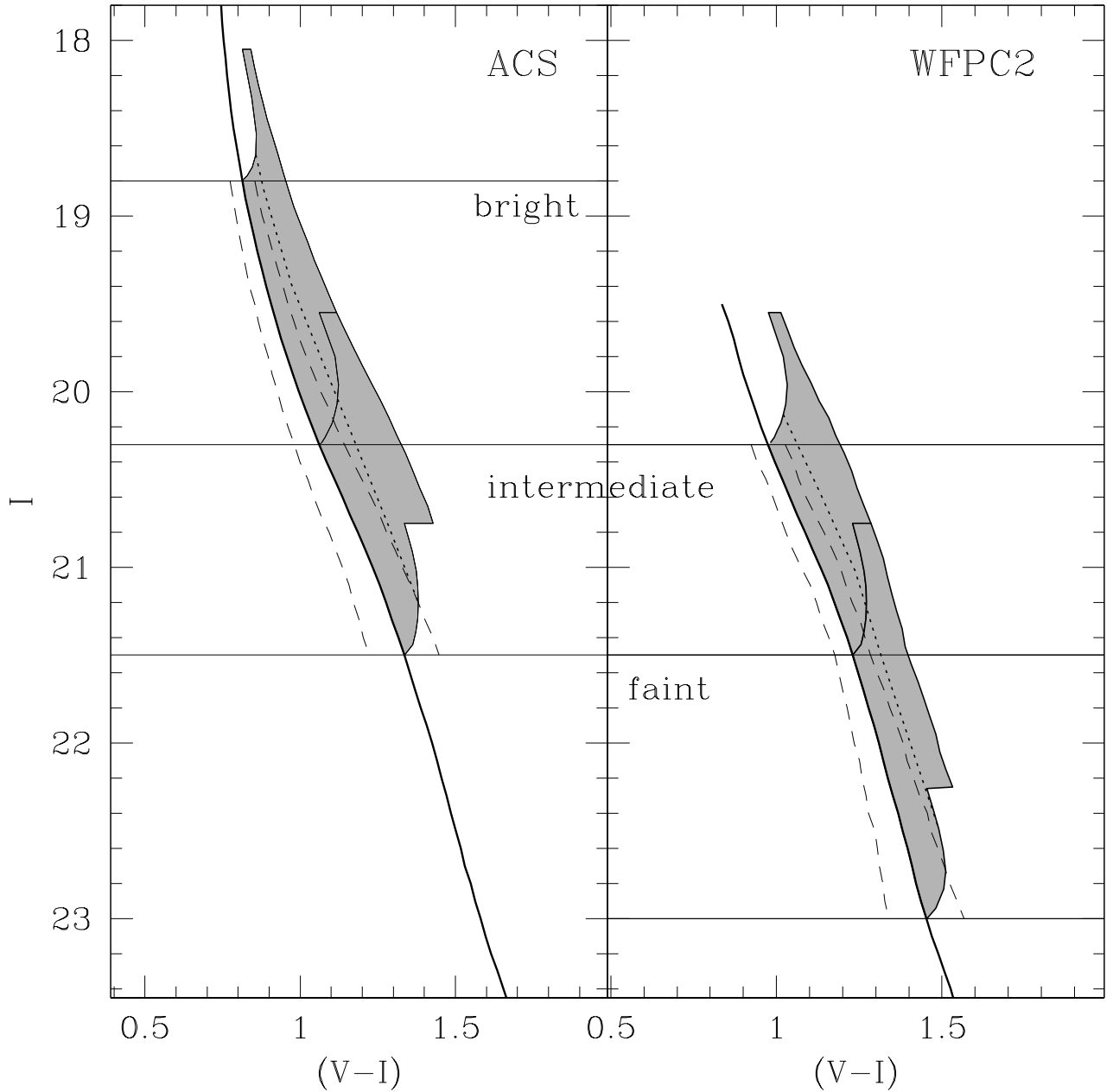}
\caption{Selection boxes used to define the ``MS population'' and the
  ``binary population'' for the two data-sets. The thick solid line
  marks the MSRL. The dashed lines bound the ``MS population'', made of
  stars with a color difference from the MSRL smaller than three times
  the typical photometric error at that magnitude level (see
  Sect. \ref{analysis}). The gray region marks the ``binary population''
  selection box, with its left-hand side corresponding to the redder
  boundary of the MS population region (right-hand dashed line) and
  its right-hand side corresponding to the equal-mass binary boundary
  shifted to the red by three times the photometric error
  (Sect. \ref{fmin}). The dotted line represents the locus defined by binary
  systems with mass-ratio $q=0.6$.}
\label{regions}
\end{center}
\end{figure}

The high photometric quality and the spatial coverage of the data-sets
previously described allowed us to study the binary fraction at
different distances from the cluster centre.  In particular, here we
have defined three concentric annuli bounded by the core radius and
the half-mass radius.  The adopted centre of gravity and structural
parameters have been recently determined from resolved star counts
(Dalessandro et al. 2011): the coordinates of the centre are
$\alpha_{\rm J2000} = 16^{\rm h}\, 57^{\rm m}\, 8.92^{\rm s}$,
$\delta_{J2000} = -4\arcdeg\,5\arcmin\, 58.07\arcsec$; the core,
half-mass and tidal radii are $r_c=48\arcsec$, $r_h= 147\arcsec$, and
$r_t=19.3\arcmin$, respectively. This center is located at $\sim3.5\arcsec$ 
North-West from the one quoted by Goldsbury et al 2010, a difference that 
has no impact on the following analysis and the obtained results. 
Hence, the first two radial bins
($r<r_c$ and $r_c<r<r_h$) are sampled by the ACS data-set, while the
third one ($r>r_h$) is covered by the WFPC2 data (see Fig. \ref{map}).
Since the two data-sets have different saturation and completeness
levels (see Sect. \ref{data}), we perfomed the analysis in two
different magnitude ranges: the adopted cuts along the MSRL are
$18.8<I<21.5$ for the ACS sample, and $20.3<I<23$ for the WFPC2 one
(see Figs. \ref{cmd} and \ref{regions}).  These intervals define what
we call the ``\emph{full} magnitude range'' of the two data-sets. Then,
with the aim of having an interval of magnitudes in common between the
two samples where to directly compare the computed binary fractions,
we considered three magnitude sub-ranges defined as follows: a
``\emph{bright} range'' corresponding to $18.8<I<20.3$, an
``\emph{intermediate} range'' at $20.3<I<21.5$, and a ``\emph{faint}
range'' at $21.5<I<23$ (all the quoted magnitude values are measured
along the MSRL). As is apparent from Figs. \ref{cmd} and \ref{regions},
the \emph{bright} range is probed only by the ACS data-set, the
\emph{faint} range is found only in the WFPC2 sample, while the
\emph{intermediate} range is in common between the two.

\subsection{The minimum binary fraction}\label{fmin}

We first estimated the \emph{minimum binary fraction} ($\xi_{min}$),
which is the fraction of binary systems with a mass ratio $q_{min}$
large enough to make them clearly distinguishable from the single-star
MS. It is clear that the value of $q_{min}$ depends directly on the
photometric errors and $\xi_{min}$ represents only a sub-sample of the
whole population of binaries, but it has the advantage of being a
purely observational quantity.  In this case we define the
\emph{``binary population''} as the set of stars located in the CMD
between the following boundaries (see gray region in
Fig. \ref{regions}): the left-hand boundary is the line corresponding
to a color difference from the MSRL equal to three times the
photometric error at any magnitude level (right dashed line); the
right-hand boundary is the line at a color difference from the
equal-mass binary sequence equal to three times the photometric error;
the upper and lower boundaries are set by the largest and the smallest
primary mass (corresponding to the quoted bright and faint cuts of the
various magnitude ranges along the MSRL), combined with all the
possible mass ratios. In other words, the ``binary population'' includes
all binary systems with primary mass set by the considered magnitude
ranges and with $q_{min}\le q\le 1$, also taking into account the
effect of photometric errors.

For each of the considered radial bins and magnitude ranges we
estimated the minimum binary fraction by performing all the steps
described in Sect. \ref{analysis} and, in much more detail, in S07.
The results are presented in Table \ref{tab:fmin}. As is apparent,
$\xi_{min}$ monotonically decreases from the center to the outskirts,
in agreement with previous findings and with theoretical
predictions (see Sect. \ref{discuss}). In the \emph{full} magnitude
range, such a radial variation ranges from $\sim 6\%$ at $r<r_c$,
to $\sim 1\%$ at $r>r_h$. There also seems to be a trend with 
magnitude, especially in the central bin, where $\xi_{min}$ varies
from $\sim 8\%$ in the \emph{bright} range, to $\sim 5\%$ in the
\emph{intermediate} one.

However, since the photometric error depends on magnitude, the
value of $q_{min}$ changes in the considered luminosity ranges: for
decreasing luminosity, $q_{min}$ varies from 0.5 to 0.6 in both the
ACS and the WFPC2 samples.  Hence, the derived values of $\xi_{min}$
are neither strictly comparable to one other, nor to the
estimates presented in different works.  We have therefore computed
the fraction of binaries with mass ratios larger than a fixed value
$q=0.6$ ($\xi_{\geq0.6}$). This value has been chosen as a compromise between having
enough statistics and avoiding contamination from single stars
(indeed, the line corresponding to $q=0.6$ in the CMD always runs to
the right-hand side of the MS population boundary).  In this case the
``binary population'' is made up of stars that, in the CMD, are located
between the line of constant $q=0.6$ (left boundary; see dotted lines
in Fig. \ref{regions}) and the right-hand boundary defined above. Its
ratio with respect to the total number of stars gives the fraction of
binaries with $q\ge0.6$, which is presented in Table
\ref{tab:fq06}. Obviously, the obtained values are smaller than the
corresponding minimum fractions $\xi_{min}$ in Table
\ref{tab:fmin}. We also note that the same behaviors discussed above
are still present, thus again suggesting that the trend with
magnitude could be real. One possible explanation for the trend with 
magnitude could be that
\emph{bright} range systematically samples more massive stars, which
are also expected to be more centrally segregated.

\subsection{The global binary fraction} \label{fglob}
In order to estimate the overall binary content of M10, independently
of the value of $q$, we also computed the \emph{global binary
  fraction} ($\xi_{TOT}$).  This requires us to perform simulations of
single and binary star populations assuming different input values of
the global binary fraction ($\xi_{\rm in}$) and then determining
$\xi_{TOT}$ from the comparison between the artificial and the
observed CMDs: the value of $\xi_{\rm in}$ that provides the best
match between the two CMDs is adopted as the global binary fraction
$\xi_{TOT}$ (see Bellazzini et al. 2002 and S07 for a detailed
description of the procedure).

\begin{figure}
\begin{center}
\includegraphics[scale=0.9]{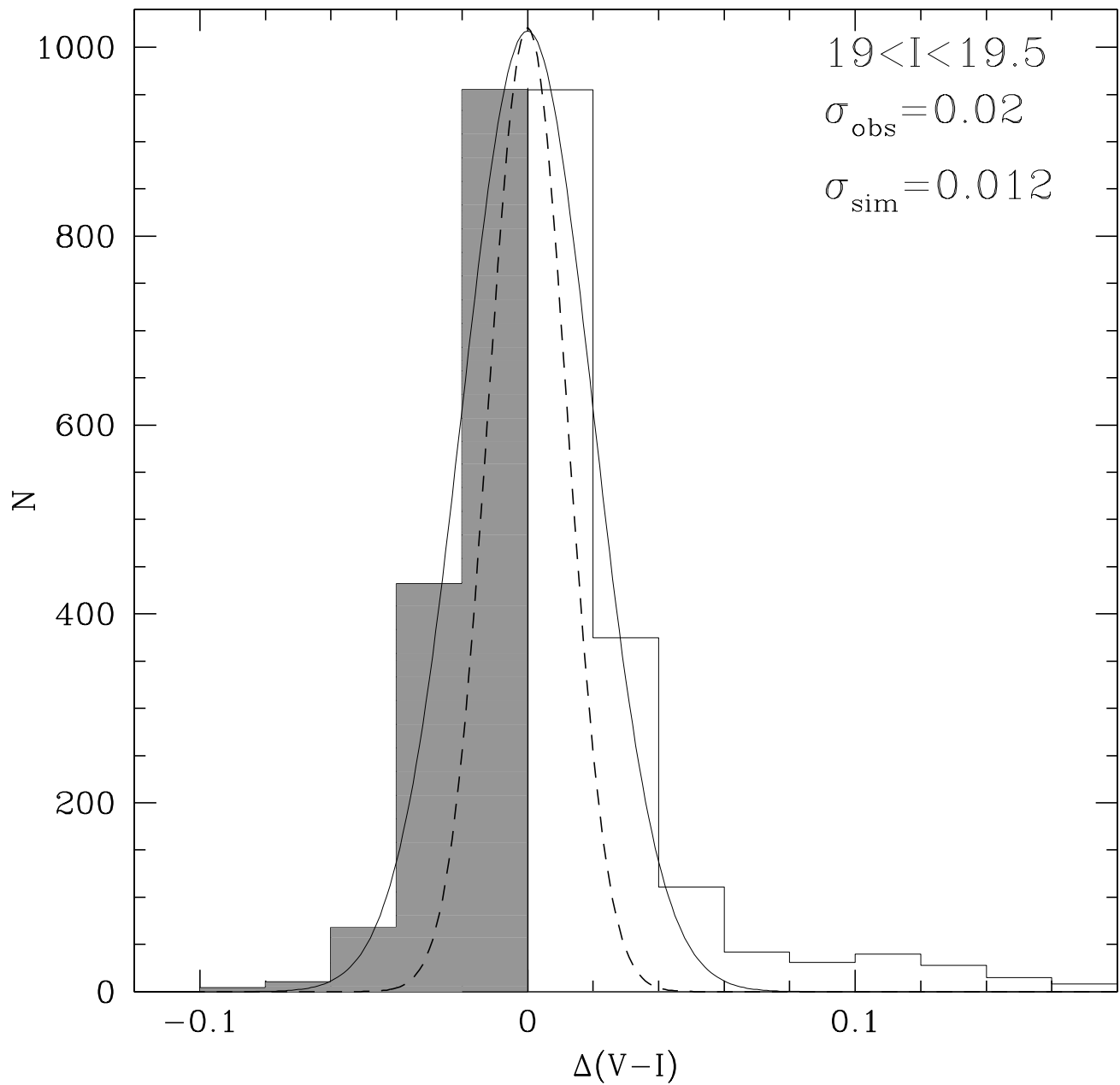}
\caption{Observed color distribution of MS stars with respect to the
  MSRL, in the magnitude range $19<I<19.5$ for the ACS data-set
  (histogram).  The solid line corresponds to the Gaussian that
  best-fits the blue-side of the observed distribution (gray
  histogram), while the red-side has not be taken into account since
  it also includes the contribution of binaries. The dashed line is a
  Gaussian with a dispersion equal to the formal photometric error
  derived from the artificial star simulations (B10).}
\label{photerr}
\end{center}
\end{figure}

\begin{figure}
\begin{center}
\includegraphics[scale=0.9]{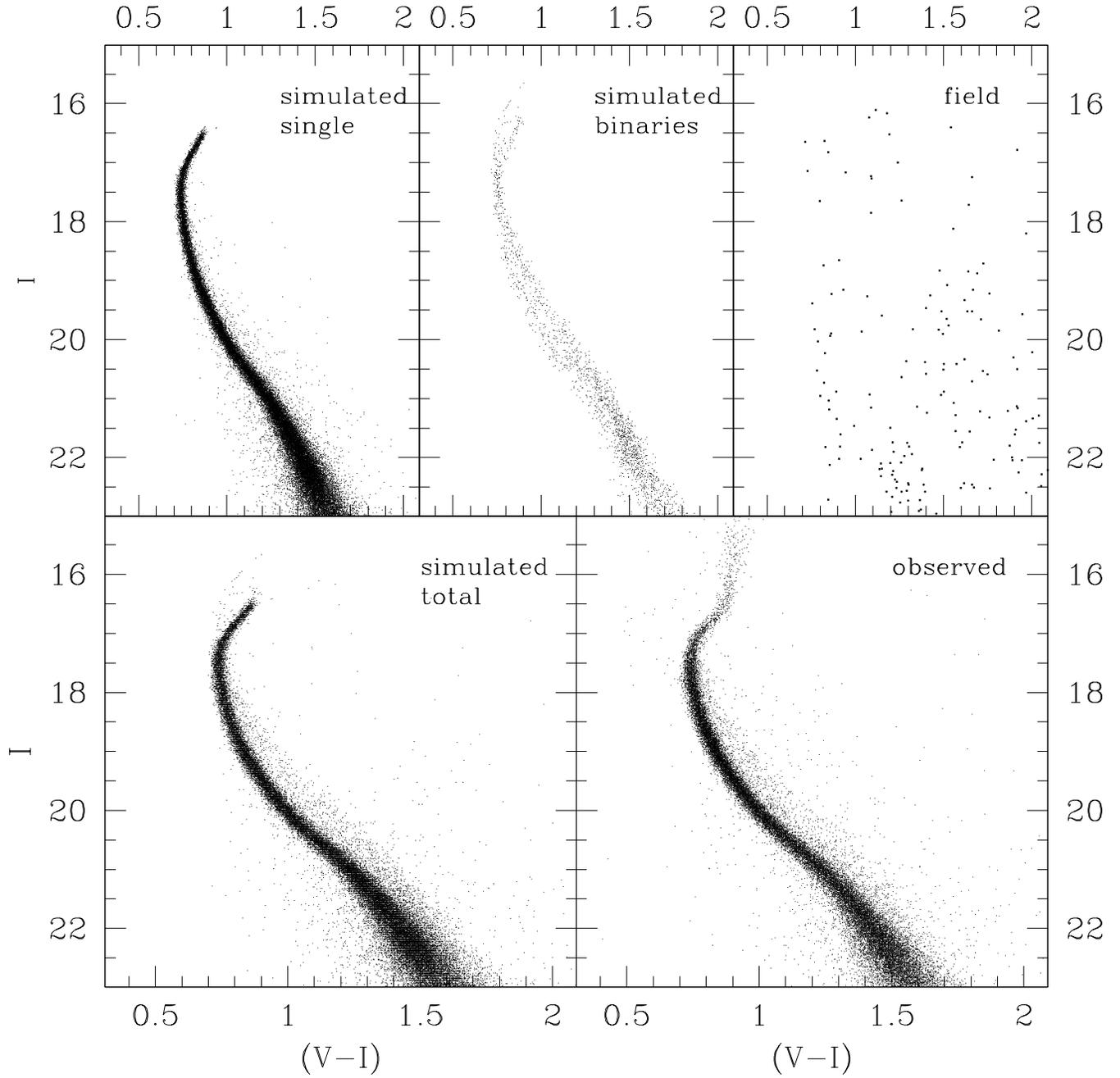}
\caption{CMDs corresponding to the simulated single MS stars,
  binaries, and field stars for the case of the ACS sample,
  $r_c<r<_h$, and $\xi_{\rm in}=6.3$ (upper panels). The lower panels
  show the comparison between the combined synthetic CMD (left-hand
  side) and the observed one (right-hand side).}
\label{simu}
\end{center}
\end{figure}

Once we assumed an input value of the binary fraction ($\xi_{\rm in}$),
for each of the considered radial and magnitude bins we have built a
sample of $N_{\rm MS}$ and $N_{bin}$ stars, with $N_{bin}= N
\xi_{\rm in}$, $N$ being the number of observed objects (after having
taken into accout the number of contaminating field stars, discussed
in Sect. \ref{analysis}) in that bin, and $N_{\rm MS}$ being $N(1-\xi_{min})$.
The MS stars have been simulated by
randomly extracting $N_{\rm MS}$ values of the mass from the
present-day cluster mass function derived by B10\footnote{B10
  suggested that for stars below $0.5 M_\odot$, the slope of the mass
  function decreases from $0.23$ to $-0.83$ moving from the inner to
  the outer regions (for reference, the slope of the canonical
  Salpeter mass function would be $-2.35$).  In order to understand
  how the assumed mass function may affect the binary fraction
  estimates, we have re-computed $\xi_{TOT}$ by adopting the core mass
  function for the whole cluster.  Within the errors, the resulting
  values of $\xi_{TOT}$ turn out to be in agreement with those
  presented in Table \ref{tab:fglob}, thus guaranteeing that the
  global binary fraction is just mildly sensitive to changes in the
  mass function.}, and transforming the masses into luminosities by
using the \citet{baraf97} isochrones.  Then, from the artificial-star
catalogue previously described, we have randomly selected an object
with similar ($\Delta I<0.1$) magnitude and, if recovered, we assigned
its output $I$ and $V$ magnitudes to the considered MS star. In order
to simulate the binary systems we randomly extracted $N_{bin}$ values
of the mass of the primary component from the \citet{kroupa02} initial
mass function, and $N_{bin}$ values of the binary mass ratio from the
$f(q)$ distribution observed by \citet{fisher05} in the solar
neighborhood, thus also obtaining the mass of the secondary. After
transforming masses into luminosities and summing up the fluxes of the
two components, an object with similar magnitude was randomly
extracted from the artificial-star catalogue and, if recovered by the
photometric analysis, the shifts between its input and output
magnitudes were assigned to the considered binary system. Finally, the
field stars were added to the sample.  The result of this procedure is
a list of synthetic stars with the same characteristics of real stars
and containing a given fraction of binaries ($\xi_{\rm in}$).  To be
precise, the MSs of the resulting artificial CMDs are narrower than
the observed ones, because the formal photometric errors of the
artificial star catalogue systematically underestimate the true
observational uncertainties.  This is apparent in Figure
\ref{photerr}, where, for the magnitude range $19<I<19.5$, the
histogram corresponds to the distribution of the observed color
differences $\Delta(V-I)$ with respect to the MSRL, the solid line is
the best-fitting Gaussian of the blue-side of this distribution (gray
histogram; the red-side has been ignored because it also includes the
contribution of binaries and blends), and the dashed line is a
Gaussian with a dispersion obtained by adopting the formal photometric
error of the artificial star catalogue.  In order to correct for this
bias and adopt realistic values of the photometric uncertainty, we
increased the formal errors $\sigma_I$ and $\sigma_V$ thus to
reproduce the observed error distribution as a function of
magnitude. As a check, we verified that the width of the resulting
color distribution with respect to the MSRL well matches the observed
one.  An example of the synthetic CMD thus obtained, compared to the
observed one is shown in Fig. \ref{simu}.  From the simulated
catalogue we then computed the ratio $r_{sim}=N_{bin}^{\rm
  sim}/N_{\rm MS}^{\rm sim}$ between the number of synthetic stars
belonging to the ``binary population'' defined in Sect. \ref{fmin}, and
that of the synthetic ``MS population''.  The same was done for the
observed data-sets, thus obtaining $r_{obs}=N_{bin}^{\rm obs}/N_{\rm
  MS}^{\rm obs}$.

\begin{figure}
\begin{center}
\includegraphics[scale=0.9]{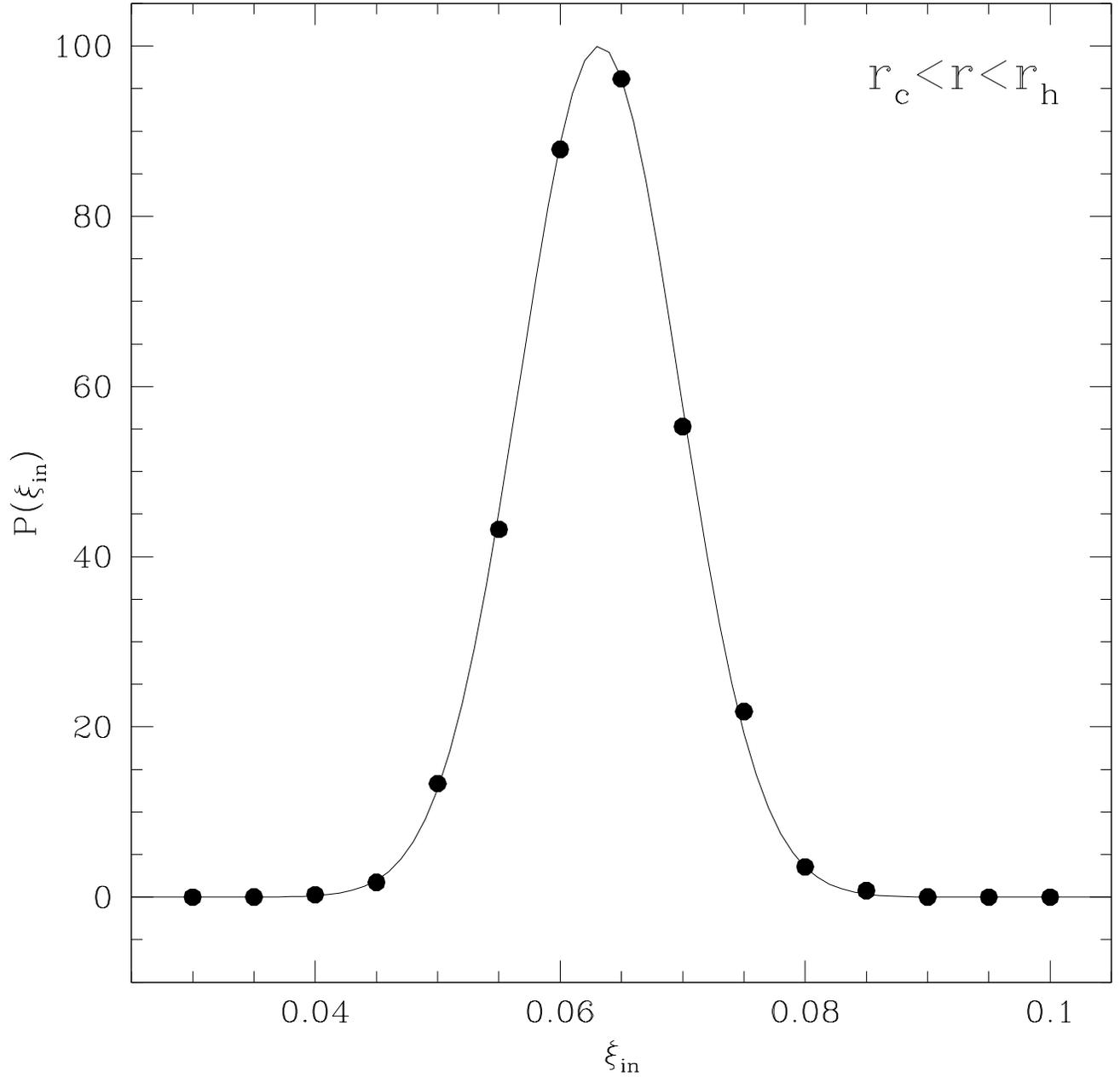}
\caption{Probability distribution of the adopted input binary
  fractions $\xi_{\rm in}$, for the case of the ACS sample,
  $r_c<r<r_h$ and the \emph{intermediate} magnitude range.  The mean
  and the dispersion of the best-fitting Gaussian give the global
  binary fraction and its error: $\xi_{TOT}=(6.3\pm0.6)\%$ (see Table
  \ref{tab:fglob}).}
\label{prob}
\end{center}
\end{figure}

For every value of $\xi_{\rm in}$, from $0.5\%$ to $25\%$ with steps 
of $0.5\%$,  the entire procedure was
repeated 100 times. Then, the penalty function $\chi^2(\xi_{\rm in})$
was computed as the summation of $(r_{sim, i} -r_{obs})^2$ for
$i=1,100$, and the associated probability $P(\xi_{\rm in})$ was derived. 
To illustrate this, Figure \ref{prob} shows the
distribution of $P$ as a function of the adopted values of $\xi_{\rm
  in}$, ranging from 3\% to 10\%.  The mean of the best-fitting
Gaussian gives the global binary fraction ($\xi_{TOT}$) and its
dispersion has been adopted as the error.  The values of $\xi_{TOT}$
obtained in the various radial and magnitude ranges are reported in
Table \ref{tab:fglob}.  The global binary fraction shows the same
radial behavior observed for $\xi_{min}$, varying from $\sim 14\%$ or
$\sim 10\%$ in the cluster core (for the \emph{full} and the
\emph{intermediate} magnitude ranges, respectively), down to $\sim
1.5\%$ in the outskirts (for both). As before we find a dependence of
the binary fraction on the magnitude. This could be an effect of
mass segregation, since the average binary mass in the \emph{bright},
\emph{intermediate} and \emph{faint} ranges is $M\sim 1.1, 0.8, 0.5
M_\odot$, respectively.  However it could also depend on the assumed
mass-ratio distribution and the estimate of blended sources, and
future studies will be required to resolve this.

\section{DISCUSSION}\label{discuss}

We have presented a homogeneous analysis of the binary fraction in M10
as a function of the radial distance from the cluster centre, from the
core region, out to $\sim 2 r_h$. Within the errors, the derived
\emph{core} binary fraction is consistent with that measured in other
GCs, which have typical values of $\xi_{TOT}$ spanning 
from $\sim 10\%$ to $\sim 25\%$ (S07; Davis et al. 2008)
but it is significantly smaller than
that estimated for the faintest clusters in the sample of
\citet{sol07}, which reach also binary fractions 
$\xi_{TOT}\sim 50\%$. This is in agreement with the quoted anti-correlation
between binary fraction and total luminosity \citep{milone08, sol08,
  sol10}. Also the binary fraction beyond the half-mass radius ($\sim
1\%$) is consistent with previous estimates in GCs \citep[see Table
  1][]{davis08}.

\begin{figure}
\begin{center}
\includegraphics[scale=0.9]{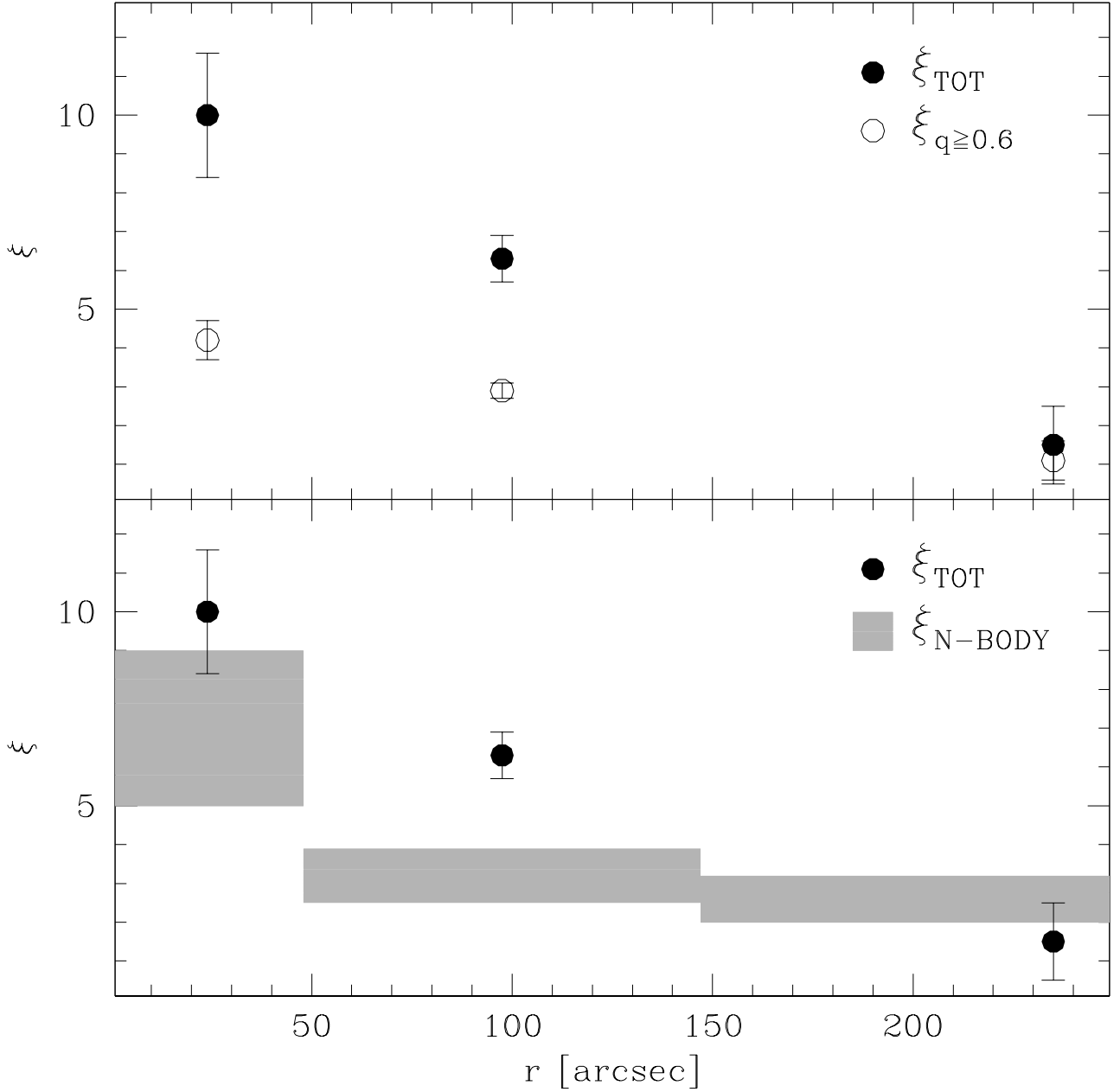}
\caption{\emph{Upper panel:} Radial behavior of the global binary
  fractions (black dots) and of the fraction of binaries with mass
  ratio $q\ge0.6$ (empty dots) estimated for the \emph{intermediate}
  magnitude range ($20.3<I<21.5$). \emph{Lower panel:} comparison
  between the observed values of $\xi_{\rm TOT}$ (the same as above;
  black dots), and the corresponding current binary fractions obtained
  from an N-body simulation that started with a 5\% primordial value
  and no central IMBH (gray regions; see text and B10).  }
\label{trend}
\end{center}
\end{figure}

The \emph{minimum} binary fraction decreases from $\sim 6\%$ within
$r_c$, to $\sim 1\%$ beyond the half-mass radius. An analogous trend
was found for the fraction of binaries with $q\ge 0.6$ and for
the \emph{global} binary fraction (Fig. \ref{trend}), the latter
varying from $\sim 14\%$ to $\sim 1.5\%$ from the core to beyond 
the half-mass radius. Such a
radial behavior is in agreement with what has been previously found in the few
other GCs where this kind of investigation has been performed
\citep[][and references in Table 1 of Davis et al. 2008]{rubai97,
  bell02, zhao05, somma09}.  It is also in agreement with the
expectations of dynamical models, where the effect is essentially 
due to the mass-segregation process, 
which leads to an increase in the number of binaries in the cluster cores
\citep[e.g.,][]{hurl07, sol08, freg09, ivanova11}.  Indeed, the
half-mass relaxation time of M10 \citep[$\sim 0.8$ Gyr, Harris 1996;
  see also][]{gnedin99,mcLvdM05} is just a small fraction ($\sim 4\%$)
of the cluster age \citep[$t\sim 13$ Gyr;][]{dot10}, so it seems safe 
to conclude that the system has already had time to achieve equipartition.

By comparing the radial variation of the MS stellar mass function
derived from the observations, with that obtained in N-body
simulations, B10 suggested that either an IMBH or a population of
binaries should be present and act as a central energy source in M10,
supressing the mass-segregation profile. In particular, the shallow 
mass-segregation profile could be modeled 
without an IMBH only when the simulations started with a primordial binary 
fraction of about $3-5\%$. Within this framework, in Figure
\ref{trend} we compare our derived values of $\xi_{TOT}$, with
those obtained from the dynamical evolution of the 5\% primordial
binary population in the 32K particle simulation of B10.  For a proper
comparison we considered a simulation snapshot at $\sim 7$
relaxation times, and only those binaries made of two MS stars and
with the primary component in the mass range $0.44\div 0.56 M_\odot$,
corresponding to the lower and upper cuts of the \emph{intermediate}
magnitude range along the MSRL. The resulting binary fractions for the
three considered radial bins are: $\xi_{N-body}=(0.070\pm0.02),
(0.032\pm 0.007), (0.026\pm 0.006)$, from the centre to the outskirts.
It is apparent from Fig. \ref{trend} that the observed binary fraction is
larger than the simulated one, especially in the core. This indicates
that the binary content of M10 is indeed sufficient to account for the 
observed mass segregation suppression, with no need to invoke an IMBH 
as additional energy source.

\acknowledgments We thank the anonymous referee for the careful reading and 
the useful comments that improved the presentation of this work.
This research is part of the project {\it COSMIC-LAB}
funded by the {\it European Research Council} (under contract
ERC-2010-AdG-267675).  The financial contribution of {\it Istituto
  Nazionale di Astrofisica} (INAF, under contract PRIN-INAF 2008) and
the {\it Agenzia Spaziale Italiana} (under contract ASI/INAF
I/009/10/0) is also acknowledged.


\begin{table}
\begin{center}
\begin{tabular}{|c|c|c|c|c|}
\hline
           &                 &               &                   &               \\
Radial     & \emph{full}     & \emph{bright} &\emph{intermediate}&\emph{faint}   \\   
bin        &\emph{mag range} &($18.8<I<20.3$)&($20.3<I<21.5$)    &($21.5<I<23$)  \\
           &		     &               &                   &               \\  
\hline
           &		     &               &                   &               \\ 
$r<r_c$    &$(6.3\pm0.4)\%$  &$(7.6\pm0.5)\%$&$(4.6\pm0.5)\%$    &  $-$          \\ 
           &		     &               &                   &               \\ 
$r_c<r<r_h$&$(3.6\pm0.2)\%$  &$(3.9\pm0.2)\%$&$(3.1\pm0.2)\%$    &  $-$          \\ 
           &		     &               &                   &               \\ 
$r>r_h$    &$(1.2\pm0.3)\%$  &      $-$	   &$(1.5\pm0.6)\%$      &$(1.1\pm0.3)\%$\\
           &		     &               &                   &               \\ 
\hline
\end{tabular}
\end{center}
\caption{Minimum binary fraction ($\xi_{min}$) of M10 in the three
  considered radial bins and the magnitude ranges defined in a
  Sect. \ref{analysis}.}
\label{tab:fmin}
\end{table} 

\begin{table}
\begin{center}
\begin{tabular}{|c|c|c|c|c|c|}
\hline
           &                 &               &                   &		 \\
Radial     & \emph{full}     & \emph{bright} &\emph{intermediate}&\emph{faint}   \\   
bin        &\emph{mag range} &($18.8<I<20.3$)&($20.3<I<21.5$)    &($21.5<I<23$)  \\
           &		     &               &                   &		 \\  
\hline
           &		     &               &                   &		 \\ 
$r<r_c$    &$(5.2\pm0.3)\%$  &$(6.2\pm0.5)\%$&$(4.2\pm0.5)\%$    &  $-$ 	 \\ 
           &		     &               &                   &		 \\ 
$r_c<r<r_h$&$(3.0\pm0.2)\%$  &$(3.2\pm0.2)\%$&$(2.9\pm0.2)\%$    &  $-$ 	 \\ 
           &		     &               &                   &		 \\ 
$r>r_h$    &$(0.8\pm0.2)\%$  &      $-$	     &$(0.7\pm0.4)\%$    &$(1.0\pm0.3)\%$\\
           &		     &               &                   &		 \\ 
\hline
\end{tabular}
\end{center}
\caption{As in Table \ref{tab:fmin}, but for the fraction of binaries
  with mass ratio $q\ge0.6$ ($\xi_{q\ge 0.6}$).}
\label{tab:fq06}
\end{table}

\begin{table}
\begin{center}
\begin{tabular}{|c|c|c|c|c|}
\hline
           &		    &		     &  	         & 	         \\
Radial     & \emph{full}    & \emph{bright} & \emph{intermediate}&\emph{faint}   \\
bin        & \emph{range}   &($18.8<I<20.3$) &($20.3<I<21.5$)    &($21.5<I<23$)  \\  
           &		    &		     &  	         & 	         \\
\hline
           &		    &		     &  	         & 	         \\
$r<r_c$    &$(13.8\pm1.4)\%$&$(15.1\pm1.9)\%$&$(10.0\pm1.6)\%$   &  $-$	         \\
           &		    &		     &  	         & 	         \\
$r_c<r<r_h$&$(7.4\pm0.6)\%$ &$(7.6\pm0.8)\%$ &$(6.3\pm0.6)\%$    &  $-$	         \\
           &		    &		     &  	         & 	         \\
$r>r_h$    &$(1.5\pm0.6)\%$ &	   $-$       &$(1.5\pm1.0)\%$    &$(1.5\pm0.7)\%$\\
           &		    &		     &  	         & 	         \\
\hline
\end{tabular}
\end{center}
\caption{Global binary fraction ($\xi_{TOT}$) of M10 for the
  considered radial and magnitude intervals.}
\label{tab:fglob}
\end{table}

\newpage 
 
\end{document}